\documentclass[aps,preprint,showkeys,nofootinbib,superscriptaddress]{revtex4-1}

\usepackage{amsfonts}
\usepackage{amsmath}
\usepackage{subfig}
\usepackage{mathtools,slashed,tensor}
\usepackage[font=scriptsize,labelfont=bf]{caption}
\usepackage{float}

\usepackage{hyperref}
\usepackage{color}
\definecolor{darkgreen}{rgb}{0,0.35,0}

\newcommand{\ucharles}{Faculty of Mathematics and Physics, Charles University, V Hole\v{s}ovi\v{c}k\'ach 2, 18000 Prague 8, Czech Republic}

\begin{document}
\title{Generalized uncertainty principle in graphene}

\author{A.~Iorio}
\email{alfredo.iorio@mff.cuni.cz}
\affiliation{\ucharles}
\author{P.~Pais}
\email{pais@ipnp.troja.mff.cuni.cz}
\affiliation{\ucharles}

\begin{abstract}
We show that, by going beyond the low-energy approximation for which the dispersion relations of graphene are linear, the corresponding emergent
field theory is a specific generalization a Dirac field theory. The generalized Dirac Hamiltonians one obtains are those compatible with specific generalizations of the uncertainty principle.
We also briefly comment on the compatibility of the latter with noncommuting positions, $[x_i,x_j] \neq 0$, and on their possible physical realization.
\end{abstract}


\maketitle

\section{Introduction}

At low energy, the quasiparticles of graphene responsible for its transport properties, have a well know description in terms of an emergent Dirac field theory, both in flat and curved spacetimes. See \cite{grapheneqft} for the original theory, \cite{geimnovoselovDiracElectrons} for the first experiments, and the reviews \cite{pacoreview2009,grapheneqftincurvedspace}. Recently \cite{IJMPD_Paper}, the possibility that a \textit{deformed} Dirac structure could still exist was explored, in a range of energies beyond such linear regime.

Indeed, as we shall recall later, it was found in \cite{IJMPD_Paper} that a Dirac structure of a generalized kind is present, beyond the linear approximation, and it is precisely of the specific kind obtained earlier in quantum gravity scenarios with a \textit{minimal length}, see \cite{Amelino-Camelia} for a review. This is as it must be: the higher the energy (or, equivalently, the shorter the wavelength) of the $\pi$-electron quasiparticles, the more the space they see while they propagate looses its smoothness, and the effects of the discrete honeycomb lattice (see Fig.~\ref{honeycomb}) show-up\footnote{In fact, even in the continuum/large-wavelength regime, memories of the lattice are not completely gone. On this see, e.g., \cite{IorioPais}.}. This is also consistent with doubly special relativity \cite{Amelino-Camelia} (DSR), with its maximal Planck energy scale, where the smooth manifold structure of spacetime breaks down, along with the local Lorentz invariance.

These scenarios, in fundamental research, are also the stage for certain generalized uncertainty principles (GUPs) stemming from the presence in the theory of a fundamental length scale \cite{DasVagenas,AhmedAli1,AhmedAli2,AhmedAli3}. Those differ considerably from other GUPs in the literature, see, e.g. \cite{Maggiore}. Noticeably, the latter GUPs would not apply here, while the former do.

We shall also show that noncommutativity, either of coordinates or of momenta, is supported by such GUPs. In particular, we shall comment on the fact that a specific kind of $[x_i,x_j]$ (or of $[p_i,p_j]$) can be accommodated in this context, and shall point to the role of an external magnetic field for its realization.

It is important to appreciate that these features of graphene can be useful for high energy theoretical research, because there are many open questions there that could be addressed in this way. For instance, previous work \cite{weylgraphene,hawkinggrapheneplb,hawkinggrapheneprd,IORIO2018265,icrystals}, has extensively explored the possibility to realize on graphene black hole physics within the standard description of a quantum field on (nearly everywhere) continuous spacetimes. The present approach could open the way to explore what happens to such physics when the granular effects of space(time) are taken into account, see, e.g., \cite{SCARDIGLI2017242}. One key example that comes to the mind is the experimental realization of quantum or noncommutative (or both) corrections to the Bekenstein formula for the black-hole entropy.

In the following, we shall recall how the Dirac structure survives beyond the linear regime, in the hypothesis that $[x_i,x_j] = 0 = [p_i,p_j]$. We then briefly comment on the compatibility of such structures with noncommutativity, and close with some conclusions.

\section{Dirac structure beyond the linear regime} \label{GUPcommutative}

\begin{figure}
\begin{center}
\includegraphics[width=0.8 \textwidth]{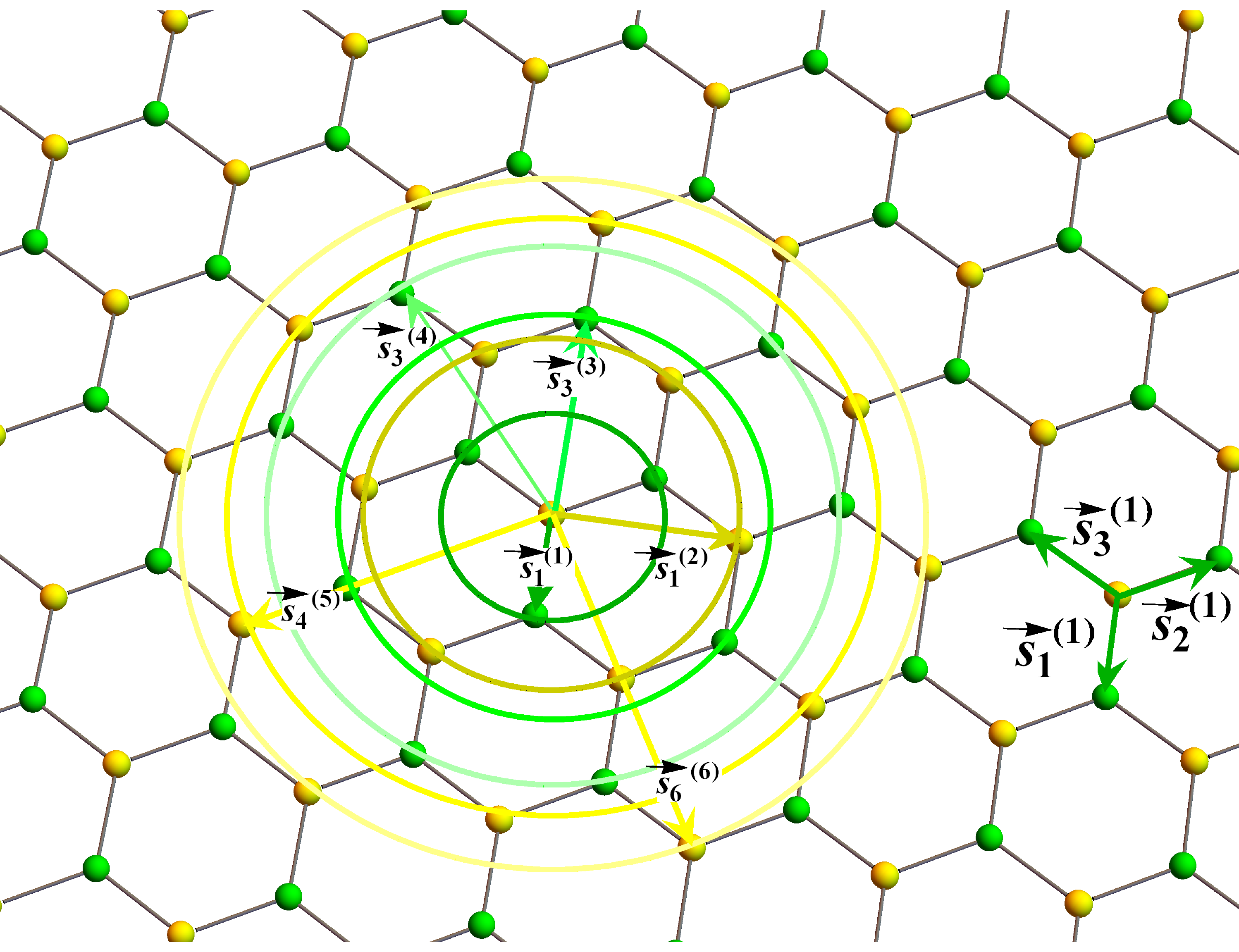}
\end{center}
\caption[3pt]{{\protect\small {The honeycomb lattice of carbons making graphene.
The two Bravais sublattices, $L_A$ and $L_B$, are identified by
different colors (green and yellow).
The three vectors connecting nearest neighbors are explicitly shown,
$\vec{s}^{(1)}_{1}= \ell(0,-1)$, $\vec{s}^{(1)}_{2}=\frac{\ell}{2}(\sqrt{3},1)$,
$\vec{s}^{(1)}_{3}=\frac{\ell}{2}(-\sqrt{3},1)$ with $\ell\approx1.42${\AA}
the carbon-to-carbon distance, whereas only one representative per $m$ of the
$\vec{s}^{(m)}_{i}$s, till $m=6$, is shown. See \cite{IJMPD_Paper}.
}}}%
\label{honeycomb}%
\end{figure}

The honeycomb lattice of graphene, see Fig.~\ref{honeycomb}, has \textit{two} atoms per unit cell. The quantum state associated to such configuration is
$|\Phi_{\vec{k}} \rangle =  a_{\vec{k}} |\phi^{A}_{\vec{k}} \rangle + b_{\vec{k}} |\phi^{B}_{\vec{k}} \rangle$. Here ${\vec{k}}$ labels the crystal quasi-momentum, $a_{\vec{k}}$ and $b_{\vec{k}}$ are complex functions, and $\phi^{I}_{\vec{k}} (\vec{r}) = \langle \vec{r}|\phi^{I}_{\vec{k}} \rangle = e^{i \vec{k} \cdot \vec{r}} \varphi^{I} (\vec{r})$, with $I = A, B$, are Block functions, with periodic orbital wave-functions $\varphi^{I} (\vec{r})$ localized at the sites of the Bravais sublattices, either $L_A$ or $L_B$.

From the Schr\"{o}dinger equation $\hat{H}_{\vec{k}} |\Phi_{\vec{k}} \rangle = E(\vec{k}) |\Phi_{\vec{k}} \rangle$, we obtain
\begin{equation}
{\cal H}_{\vec{k}} \equiv \langle \Phi_{\vec{k}}| \hat{H}_{\vec{k}} |\Phi_{\vec{k}} \rangle
   = \int d^2 {\vec{r}} d^2 {\vec{r}'} \langle \Phi_{\vec{k}} |\vec{r} \rangle \langle {\vec{r}}| \hat{H}_{\vec{k}} |\vec{r}' \rangle \langle {\vec{r}}' |\Phi_{\vec{k}} \rangle \,.
   \label{fieldHfirst}
\end{equation}
We rewrite the overall interaction as a sum of interactions between subsequent orders of neighbors, and go up to the $m^{\textrm{th}}$-near neighbors. It can be shown \cite{IJMPD_Paper} that the field Hamiltonian (\ref{fieldHfirst}) becomes
\begin{equation*}
{\cal H}_{\vec{k}} = \sum_{m \in \textrm{diag}} (\epsilon^{(0)}  \varsigma_m + \eta_m )  {\cal F}_m ({\vec{k}}) (a^*_{\vec{k}} a_{\vec{k}} + b^*_{\vec{k}} b_{\vec{k}})
+ \left( \sum_{m \in \textrm{off}} (\epsilon^{(0)}  \varsigma_m + \eta_m )  {\cal F}^*_m ({\vec{k}}) a^*_{\vec{k}} b_{\vec{k}} + h.c. \right) \;,
\end{equation*}
where $\varsigma_m$ and $\eta_m$ are the overlapping and the hopping parameters, respectively, and ${\cal F}_m ({\vec{k}}) \equiv \sum_{i=1}^{n_m} \exp \{ i {\vec{k}} \cdot {\vec{s}^{(m)}}_i \}$. We have
\begin{equation}\label{firstfunction}
  {\cal F}_1 = \sum_{i= 1}^{3} e^{i {\vec{k}} \cdot {\vec{s}^{(1)}}_i} =
  e^{- i \ell k_y} [1 + 2 e^{i \frac{3}{2} \ell k_y} \cos(\frac{\sqrt{3}}{2} \ell k_x)] \,,
\end{equation}
where $\ell$ is the lattice spacing, and the important relation ${\cal F}_2 = |{\cal F}_1|^2 - 3$, holds \cite{IJMPD_Paper}.

By setting the zero of $E(\vec{k})$ to $\epsilon^{(0)}$, the Schr\"{o}dinger equation becomes
$\psi^\dag_{\vec{k}} \widehat{{\cal H}}_{\vec{k}} \psi_{\vec{k}} = E(\vec{k}) \psi^\dag_{\vec{k}} \widehat{{\cal S}}_{\vec{k}} \psi_{\vec{k}}$
where $\psi^\dag_{\vec{k}} \equiv (a^*_{\vec{k}}, b^*_{\vec{k}})$, and the Hamiltonian matrix is
$\widehat{{\cal H}}_{\vec{k}} = h_{\vec{k}}^{\textrm{diag}} \mathbf{{1}}_2 + \textrm{Re} h_{\vec{k}}^{\textrm{off}} \sigma_x
                      + \textrm{Im} h_{\vec{k}}^{\textrm{off}} \sigma_y$, while the overlapping matrix is
$\widehat{{\cal S}}_{\vec{k}} = {\cal S}_{\vec{k}}^{\textrm{diag}} \mathbf{{1}}_2 + \textrm{Re} {\cal S}_{\vec{k}}^{\textrm{off}} \sigma_x
                      + \textrm{Im} {\cal S}_{\vec{k}}^{\textrm{off}} \sigma_y$, with $\sigma_x$ and $\sigma_y$ the standard Pauli matrices.
Evidently, it is the concurrence of two atoms per unit cell, and of the Hermiticity of the Hamiltonian, that are behind the emergence of the Pauli matrices, although this is not enough to have a Dirac structure, as we need to focus on the two solutions, $E_\pm$, to the secular equation
${\rm det}   \left( \widehat{{\cal H}}_{\vec{k}} - E  \widehat{{\cal S}}_{\vec{k}} \right) = 0$.

This gives an involved expression in terms of the ${\cal F}_m$s, and in general there is no Dirac structure behind such dispersion relations, but the occurrence that ${\cal F}_2 = |{\cal F}_1|^2 - 3$, allows us to write $E_\pm = \eta_1 \left( \pm |{\cal F}_1| - \tilde{A}  |{\cal F}_1|^2 \right)$, where $\tilde{A} = {\epsilon'}_1/ \eta_1$ is a dimensionless parameter\footnote{Here we are entitled to consider one  Dirac point only \cite{IORIO2018265}, we shift the zero of the energy again, $\epsilon_1 \to {\epsilon'}_1 \equiv \epsilon_1 - \epsilon_0$, and we drop terms of the kind $O(\varsigma_1^2)$ and $O(\varsigma_1 \eta_2)$ or higher.}. By defining ${\vec{P}} \equiv (\hbar / \ell)  \left( Re {\cal F}_1, Im {\cal F}_1  \right)$, $A \equiv (\ell / \hbar ) \tilde{A}$, and\footnote{Notice that the Fermi velocity is usually defined as $v_F = 3/2 \eta_1 \ell / \hbar$, that is $v_F = 3/2 V_F$. See \cite{pacoreview2009}.} $V_F \equiv \eta_1 \ell / \hbar$ we arrive at
\begin{eqnarray}
  E_\pm  \equiv & V_F \left( \pm |\vec{P}| - A \; |\vec{P}|^2 \right) \,. \label{DispRelLargeP}
\end{eqnarray}
This, including the important minus sign in front of $A$, is precisely what obtained in \cite{AhmedAli1,AhmedAli2,AhmedAli3} when generalizing the Dirac dispersion relations, to accommodate the effects of a GUP with a minimal fundamental length $A \propto \ell$, the honeycomb lattice spacing here.

By introducing a new generalized momentum ${\vec{Q}} \equiv {\vec{P}} (1 - A |{\vec{P}}|)$, of which the former, $\vec{P}$, is the low-energy approximation, we have the Dirac Hamiltonian
\begin{equation}\label{newDiracHwithQ}
H (Q) = V_F \sum_{\vec{k}} \psi^\dag_{\vec{k}} \not\!Q \psi_{\vec{k}} \,,
\end{equation}
where we used the Pauli matrices properties to write $\not\!P \not\!P =  |\vec{P}|^2 \mathbf{1}_2$. When
\begin{equation}\label{commutativexandp}
\left[X^P_{i},X^P_{j}\right] = 0 = \left[P_{i},P_{j}\right] \,, \quad {\rm and} \quad \left[X^P_{i}, P_{j}\right] = i \hbar \delta_{i j} \,,
\end{equation}
then
\begin{equation}\label{QGUP}
\left[X^P_{i},Q_{j}\right] = i\hbar \left[ \delta_{ij} - A \left( Q \delta_{ij} + \frac{Q_i Q_j}{Q} \right)\right]
\end{equation}
that is precisely the GUP of \cite{AhmedAli1,AhmedAli2,AhmedAli3}. As long those phase space variables are concerned, this is it. Nonetheless, it is crucial to notice that here we have \textit{three} regimes, with a different kind of momentum at any given regime: two generalized momenta, $\vec{Q}$ and $\vec{P}$ (with $\vec{P}$ low-energy with respect to $\vec{Q}$), and the standard momentum $\vec{p}$ (that is low-energy with respect to $\vec{P}$, hence with respect to $\vec{Q}$).

At the first level of the ladder, $(X_i^Q, Q_j)$, full Lorentz symmetry is preserved, see (\ref{newDiracHwithQ}). At the second level, $(X_i^P, Q_j)$, the Hamiltonian preserves rotation symmetry, $SO(2)$ in this case, but full Lorentz is gone (see $Q_i Q_j / Q$ in (\ref{QGUP})). At the last level in the ladder, $(x_i, Q_j)$, not even rotation symmetry of the Hamiltonian survives (see the so-called ''trigonal warping'' \cite{pacoreview2009}, that here we naturally reproduce).

Therefore, in the hypothesis that $\left[x_i, x_j\right] = 0 = \left[p_i, p_j\right]$, and since the low energy momenta are the $p_i$s (i.e., $\left[x_{i},p_{j}\right]=i\delta_{ij}$), by expanding $\vec{P}(\vec{p})$ to $O(\ell)$ we have
\begin{eqnarray}\label{modified_commutator}
\left[x,Q_{x}\right] & = & i \left(1 - \frac{\ell}{2} Q_{x}-\frac{\ell{\epsilon'}_1}{ \eta_1}\left(|\vec{Q}|+\frac{Q_{x}^{2}}{|\vec{Q}|}\right)\right) \; , \nonumber \\
\left[x,Q_{y}\right] & = & i \left( \frac{\ell}{2} Q_{y}-\frac{\ell{\epsilon'}_1}{ \eta_1|\vec{Q}|}Q_{x}Q_{y}\right)\; , \nonumber  \\
\left[y,Q_{x}\right] & = &  i \left( \frac{\ell}{2} Q_{y}-\frac{\ell{\epsilon'}_1}{ \eta_1|\vec{Q}|}Q_{x}Q_{y}\right)   \; , \\
\left[y,Q_{y}\right] & = & i \left(1 + \frac{\ell}{2} Q_{x}-\frac{\ell{\epsilon'}_1}{ \eta_1}\left(|\vec{Q}|+\frac{Q_{y}^{2}}{|\vec{Q}|}\right)\right)\;. \nonumber
\end{eqnarray}
In \cite{IJMPD_Paper} the result up to order $O(\ell^{2})$ is given. This is a rich algebraic structure, worth investigating, especially in relation to the possibility for noncommutativity.

\section{GUP algebra and noncommutativity in graphene}\label{Section_new}

Noncommutativity of position and momentum is at the core of quantum mechanics, hence, as such, a well accepted fact of Nature. On the other hand, noncommutativity of positions among themselves still seems quite a radical and fully theoretical hypothesis.

In fact, such a possibility has been studied from many perspectives, starting from the early days of quantum mechanics with the Lorentz-symmetry-preserving model of Snyder \cite{Snyder}. It then went through different stages of fortune, from the ground-breaking mathematical framework of Connes's Noncommutative Geometry \cite{DBLP:books/daglib/0076876}, to Doplicher et al's hypothesis of a necessary outcome of the reconciliation between general relativity and quantum mechanics \cite{Doplicher1995}, to the noncommutative gauge theory as a possible low-energy limit of string theory, obtained via the Seiberg-Witten map\footnote{String theory is actually not necessary to obtain the Seiberg-Witten map, as shown by Wess and coworkers \cite{Madore2000}.} \cite{Seiberg-Witten}. The latter results, in the past decade, ignited research on the emergence of such phenomenon in condensed matter systems \cite{Jackiw2002}, especially impacts on Lorentz symmetry, e.g., in electrodynamics \cite{GURALNIK2001450, PhysRevD.69.065008}.

Let us then comment here on whether this radical idea can find some room in these settings, and what could this possibly mean from the physical point of view. For the sake of simplicity and generality of the discussion, here we do not follow too closely the notation of the previous Section, and we call the coordinates $x_i$, and the momenta $p_i$, without explicit reference to what is high-energy and what is low-energy. It should not be difficult to reconstruct the right correspondence, though.

In the previous Section, we saw  that, by considering terms beyond the linear approximation, a modified Heisenberg algebra emerges. In there, we assumed that the coordinates (and momenta) commute among themselves. Indeed, the algebra (\ref{QGUP}) is computed by employing Jacobi identities where the commutators $[x_{i},x_{j}]$ and $[p_{i},p_{j}]$ are demanded to be zero \cite{AhmedAli3}. If we allow for noncommutativity, we have many more possibilities. Here we merely explore which noncommutative algebra is compatible with the GUP emerged in graphene, see (\ref{QGUP}), up to quadratic contributions.

Suppose we have a general Weyl-Heisenberg algebra of the kind studied in \cite{AhmedAli1,AhmedAli2,AhmedAli3}
\begin{equation}\label{modifiedWeylHeisGeneral}
[x_{i},p_{j}] = i \hbar \left(\delta_{ij}\left(1+\alpha_{1}p+\beta_{1}p^{2}\right)+p_{i}p_{j}\left(\frac{\alpha_{2}}{p}+\beta_{2}\right)\right)\;,
\end{equation}
where $p\equiv\sqrt{p^2}=\sqrt{p_{i}p^{i}}$, and $\alpha$s and $\beta$s must be dimensionfull constants, $[\alpha_a] \sim 1/p$, $[\beta_a] \sim 1/p^{2}$, for $a=1,2$. Imposing the Jacobi identity,
\begin{equation}\label{Jacobi_identity}
\left[\left[x_{i},x_{j}\right],p_{k}\right]+\left[\left[x_{j},p_{k}\right],x_{i}\right]+\left[\left[p_{k},x_{i}\right],x_{j}\right]=0 \;,
\end{equation}
leads to the relation \cite{AhmedAli3}
\begin{equation}\label{Jacobi_result}
-\left[\left[x_{i},x_{j}\right],p_{k}\right]=\left( (\alpha_{1}-\alpha_{2})p^{-1}+(\alpha_{1}^{2}+2\beta_{1}-\beta_{2}) \right) \Delta_{ijk} \;,
\end{equation}
with $\Delta_{ijk}\equiv \hbar^2 (p_{i}\delta_{jk}-p_{j}\delta_{ik})$.

Equation \eqref{Jacobi_result} can be solved by demanding both its sides to be zero, which gives \cite{AhmedAli1,AhmedAli2,AhmedAli3} $\alpha_{1}=\alpha_{2}\equiv-\alpha$ and $\beta_{2}=3\alpha^{2}$, with
$\beta_{1}=\alpha^{2}$, on the one hand, and commuting coordinates, on the other hand. The final expression for the GUP is then
\begin{equation}\label{ADV_GUP}
[x_{i},p_{j}]= i \hbar \left(\delta_{ij}-\alpha\left(p\delta_{ij}+\frac{p_{i}p_{j}}{p}\right)+\alpha^{2}(p^{2}\delta_{ij}+3p_{i}p_{j})\right)\;,
\end{equation}
that, at order $\alpha$, is what we obtain in graphene too, see (\ref{QGUP}).

Noticeably, this GUP does not rule out the possibility of a noncommutative algebra, when
\begin{equation}\label{xytheta}
[x_{i},x_{j}] = i \hbar \theta_{ij} \;,
\end{equation}
and $\theta_{ij}$ is a \textit{constant}, antisymmetric, $c-$number valued quantity. Indeed, in this case, the left side of (\ref{Jacobi_result}) is still zero, so then is the right side, and the previous argument leading to (\ref{ADV_GUP}) just goes through.

On the other hand, a more general noncommutativity needs more drastic changes. For instance, one well known model is the \textit{Lie-algebra model} \cite{Madore2000},
\begin{equation}\label{xyCijk}
[x_{i},x_{j}] = i \hbar C_{ij}^k x_{k}\;,
\end{equation}
where $C_{ij}^k$ are $c-$number valued structure constants. With a noncommutativity of the kind (\ref{xyCijk}), the left side of \eqref{Jacobi_result} is no longer zero, and one cannot find a set of constant $\alpha$s and $\beta$s satisfying the equation.

Another popular realization of noncommutativity is the $q$-deformed algebra \cite{Madore2000}, based on deformed operators $x_q$ and commutators
\begin{equation}\label{qWeylHeisenberg}
x^i_q x_q^j = \frac{1}{q} R^{ij}_{kl} x^k_q x_q^l \,,
\end{equation}
with $q(\hbar)$, and $R^{ij}_{kl}$ a $c-$number valued quantity. This approach lead, for instance, to the interesting \textit{twisted coproduct} model of Balachandran et al \cite{BALACHANDRAN2006434}. For this kind of noncommutativity based on $q$-deformation, all the commutators need be, at least in principle, modified. Hence, to study whether the latter approach can find room in the graphene scenario, an analysis deeper than the one carried on here is necessary.

Similarly, one can work-out from the Jacobi for $[[p_{i},p_{j}],x_{k}]$, that noncommutativity of the momenta is also allowed along a similar pattern as for coordinates.

Therefore, we conclude that there is indeed room for noncommutativity in this graphene scenario, surely in the settings provided by (\ref{xytheta}). The questions are then: What physical situation are we describing this way? What $\theta_{ij}$ corresponds to in graphene?

We shall not answer fully these questions here, but shall only point to a scenario that could be easily realized with graphene, and that is known to have noncommutativity of the kind we are looking for. The scenario is that of planar electrons in the lowest Landau level. The energy levels of such electrons, moving within an external magnetic field of strength $B$, perpendicular to the plane, are the well-known Landau levels. In graphene, Landau levels have been realized in experiments and studied, see, e.g., \cite{PhysRevLett.98.197403}.

When the magnetic field is strong the dynamics is effectively projected to the lowest such levels\footnote{This is because the Landau levels have a separation proportional to $\sqrt{B}$ or to $B$, depending on whether we are dealing with relativistic or non-relativistic fermions.}, giving rise to \cite{Jackiw2002}
\begin{equation}
[x_{i},x_{j}]= -i\hbar\frac{c}{e B}\epsilon_{ij} \;,
\end{equation}
that in our context would mean $\theta_{ij} = \frac{c}{e B}\epsilon_{ij}$.

\section{Conclusions}

We have shown that, by going beyond the linear dispersion relations of graphene, a generalized Dirac field theory survives, and it is compatible with GUPs that require a minimal fundamental length scale. This is also consistent with Amelino-Camelia's DSR, where the smooth manifold structure of spacetime breaks down, along with the local Lorentz invariance.

While it may be surprising that some sort of field theory description still holds in graphene at higher energies (a fact due to the fortuitus occurrence ${\cal F}_2 = |{\cal F}_1|^2 - 3$), the correspondence of the latter with fields in a granular spacetime scenarios is something one should expect. In fact, the shorter the wavelength of the $\pi$-electron quasiparticles, the more the granular structure of the honeycomb lattice affects their dynamics.

It is important for high energy theory to have found here these features, because there are many open theoretical questions that could be addressed this way. To cite a key one: the experimental realization of quantum corrections to the Bekenstein formula for the black-hole entropy. Many more are, of course, possible to envisage.

We also briefly commented here on the compatibility of noncommutativity with the obtained GUP. We concluded that the simplest compatibility with the GUP modified by the graphene granular structure is given by $[x_i,x_j] = \theta_{ij}$,  with $\theta_{ij}$ a constant, antisymmetric, $c-$number valued quantity. The Lie-algebra noncommutativity is not compatible, while compatibility with $q$-deformation needs further analysis. We then pointed to a possible physical set-up, where the presence of a strong perpendicular external magnetic field $B$ could create the conditions for such noncommutativity, with $\theta_{ij} = (c /eB)  \epsilon_{ij}$, by projecting the dynamics of the $\pi$-electron quasiparticles on the lowest Landau level.

Interesting would be to compare the Lorentz violation emerging in graphene in the commutative case of Section \ref{GUPcommutative} (see the violation in (\ref{QGUP}), and the even stronger violation in (\ref{modified_commutator})), with the Lorentz violation due to noncommutativity in the approach of \cite{PhysRevD.77.048701,doi:10.1142/S0217751X02009874,ALVAREZGAUME2003293}.

\section*{Acknowledgments}

The authors are indebted to Mir Faizal, Ahmed Ali and Ibrahim Elmashad for their collaboration on these exciting matters, and thank Gaetano Lambiase and Fabio Scardigli for lively and informative discussions. P.~P. acknowledges the warm hospitality of the Centro de Estudios Cient\'ificos (CECs) in Valdivia, Chile, at the final stage of this work.

\bibliographystyle{apsrev4-1}
\bibliography{proceedings_biblio}

\begin{thebibliography}{31}%
\makeatletter
\providecommand \@ifxundefined [1]{%
 \@ifx{#1\undefined}
}%
\providecommand \@ifnum [1]{%
 \ifnum #1\expandafter \@firstoftwo
 \else \expandafter \@secondoftwo
 \fi
}%
\providecommand \@ifx [1]{%
 \ifx #1\expandafter \@firstoftwo
 \else \expandafter \@secondoftwo
 \fi
}%
\providecommand \natexlab [1]{#1}%
\providecommand \enquote  [1]{``#1''}%
\providecommand \bibnamefont  [1]{#1}%
\providecommand \bibfnamefont [1]{#1}%
\providecommand \citenamefont [1]{#1}%
\providecommand \href@noop [0]{\@secondoftwo}%
\providecommand \href [0]{\begingroup \@sanitize@url \@href}%
\providecommand \@href[1]{\@@startlink{#1}\@@href}%
\providecommand \@@href[1]{\endgroup#1\@@endlink}%
\providecommand \@sanitize@url [0]{\catcode `\\12\catcode `\$12\catcode
  `\&12\catcode `\#12\catcode `\^12\catcode `\_12\catcode `\%12\relax}%
\providecommand \@@startlink[1]{}%
\providecommand \@@endlink[0]{}%
\providecommand \url  [0]{\begingroup\@sanitize@url \@url }%
\providecommand \@url [1]{\endgroup\@href {#1}{\urlprefix }}%
\providecommand \urlprefix  [0]{URL }%
\providecommand \Eprint [0]{\href }%
\providecommand \doibase [0]{http://dx.doi.org/}%
\providecommand \selectlanguage [0]{\@gobble}%
\providecommand \bibinfo  [0]{\@secondoftwo}%
\providecommand \bibfield  [0]{\@secondoftwo}%
\providecommand \translation [1]{[#1]}%
\providecommand \BibitemOpen [0]{}%
\providecommand \bibitemStop [0]{}%
\providecommand \bibitemNoStop [0]{.\EOS\space}%
\providecommand \EOS [0]{\spacefactor3000\relax}%
\providecommand \BibitemShut  [1]{\csname bibitem#1\endcsname}%
\let\auto@bib@innerbib\@empty
\bibitem [{\citenamefont {Semenoff}(1984)}]{grapheneqft}%
  \BibitemOpen
  \bibfield  {author} {\bibinfo {author} {\bibfnamefont {G.~W.}\ \bibnamefont
  {Semenoff}},\ }\href@noop {} {\bibfield  {journal} {\bibinfo  {journal}
  {Phys. Rev. Lett.}\ }\textbf {\bibinfo {volume} {53}},\ \bibinfo {pages}
  {2449} (\bibinfo {year} {1984})}\BibitemShut {NoStop}%
\bibitem [{\citenamefont {Novoselov}\ \emph {et~al.}(2005)\citenamefont
  {Novoselov}, \citenamefont {Geim}, \citenamefont {Morozov}, \citenamefont
  {Jiang}, \citenamefont {Katsnelson}, \citenamefont {Grigorieva},
  \citenamefont {Dubonos},\ and\ \citenamefont
  {Firsov}}]{geimnovoselovDiracElectrons}%
  \BibitemOpen
  \bibfield  {author} {\bibinfo {author} {\bibfnamefont {K.~S.}\ \bibnamefont
  {Novoselov}}, \bibinfo {author} {\bibfnamefont {A.~K.}\ \bibnamefont {Geim}},
  \bibinfo {author} {\bibfnamefont {S.~V.}\ \bibnamefont {Morozov}}, \bibinfo
  {author} {\bibfnamefont {D.}~\bibnamefont {Jiang}}, \bibinfo {author}
  {\bibfnamefont {M.~I.}\ \bibnamefont {Katsnelson}}, \bibinfo {author}
  {\bibfnamefont {I.~V.}\ \bibnamefont {Grigorieva}}, \bibinfo {author}
  {\bibfnamefont {S.~V.}\ \bibnamefont {Dubonos}}, \ and\ \bibinfo {author}
  {\bibfnamefont {A.~A.}\ \bibnamefont {Firsov}},\ }\href@noop {} {\bibfield
  {journal} {\bibinfo  {journal} {Nature}\ }\textbf {\bibinfo {volume} {438}},\
  \bibinfo {pages} {197} (\bibinfo {year} {2005})},\ \Eprint
  {http://arxiv.org/abs/cond-mat/0509330} {arXiv:cond-mat/0509330
  [cond-mat.mes-hall]} \BibitemShut {NoStop}%
\bibitem [{\citenamefont {Castro~Neto}\ \emph {et~al.}(2009)\citenamefont
  {Castro~Neto}, \citenamefont {Guinea}, \citenamefont {Peres}, \citenamefont
  {Novoselov},\ and\ \citenamefont {Geim}}]{pacoreview2009}%
  \BibitemOpen
  \bibfield  {author} {\bibinfo {author} {\bibfnamefont {A.~H.}\ \bibnamefont
  {Castro~Neto}}, \bibinfo {author} {\bibfnamefont {F.}~\bibnamefont {Guinea}},
  \bibinfo {author} {\bibfnamefont {N.~M.~R.}\ \bibnamefont {Peres}}, \bibinfo
  {author} {\bibfnamefont {K.~S.}\ \bibnamefont {Novoselov}}, \ and\ \bibinfo
  {author} {\bibfnamefont {A.~K.}\ \bibnamefont {Geim}},\ }\href@noop {}
  {\bibfield  {journal} {\bibinfo  {journal} {Rev. Mod. Phys.}\ }\textbf
  {\bibinfo {volume} {81}},\ \bibinfo {pages} {109} (\bibinfo {year} {2009})},\
  \Eprint {http://arxiv.org/abs/cond-mat/0709.1163} {arXiv:cond-mat/0709.1163
  [cond-mat.mes-hall]} \BibitemShut {NoStop}%
\bibitem [{\citenamefont {Iorio}(2015)}]{grapheneqftincurvedspace}%
  \BibitemOpen
  \bibfield  {author} {\bibinfo {author} {\bibfnamefont {A.}~\bibnamefont
  {Iorio}},\ }\href@noop {} {\bibfield  {journal} {\bibinfo  {journal} {Int. J.
  Mod. Phys.}\ }\textbf {\bibinfo {volume} {D24}},\ \bibinfo {pages} {1530013}
  (\bibinfo {year} {2015})},\ \Eprint {http://arxiv.org/abs/hep-th/1412.4554}
  {hep-th/1412.4554} \BibitemShut {NoStop}%
\bibitem [{\citenamefont {Iorio}\ \emph {et~al.}(2018)\citenamefont {Iorio},
  \citenamefont {Pais}, \citenamefont {Elmashad}, \citenamefont {Ali},
  \citenamefont {Faizal},\ and\ \citenamefont {Abou-Salem}}]{IJMPD_Paper}%
  \BibitemOpen
  \bibfield  {author} {\bibinfo {author} {\bibfnamefont {A.}~\bibnamefont
  {Iorio}}, \bibinfo {author} {\bibfnamefont {P.}~\bibnamefont {Pais}},
  \bibinfo {author} {\bibfnamefont {I.~A.}\ \bibnamefont {Elmashad}}, \bibinfo
  {author} {\bibfnamefont {A.~F.}\ \bibnamefont {Ali}}, \bibinfo {author}
  {\bibfnamefont {M.}~\bibnamefont {Faizal}}, \ and\ \bibinfo {author}
  {\bibfnamefont {L.~I.}\ \bibnamefont {Abou-Salem}},\ }\href@noop {}
  {\bibfield  {journal} {\bibinfo  {journal} {Int. J. Mod. Phys.}\ }\textbf
  {\bibinfo {volume} {D27}},\ \bibinfo {pages} {1850080} (\bibinfo {year}
  {2018})},\ \Eprint {http://arxiv.org/abs/physics.gen-ph/1706.01332}
  {physics.gen-ph/1706.01332} \BibitemShut {NoStop}%
\bibitem [{\citenamefont {Amelino-Camelia}(2013)}]{Amelino-Camelia}%
  \BibitemOpen
  \bibfield  {author} {\bibinfo {author} {\bibfnamefont {G.}~\bibnamefont
  {Amelino-Camelia}},\ }\href@noop {} {\bibfield  {journal} {\bibinfo
  {journal} {Living Rev. Relat.}\ }\textbf {\bibinfo {volume} {16}},\ \bibinfo
  {pages} {5} (\bibinfo {year} {2013})}\BibitemShut {NoStop}%
\bibitem [{\citenamefont {Iorio}\ and\ \citenamefont {Pais}(2015)}]{IorioPais}%
  \BibitemOpen
  \bibfield  {author} {\bibinfo {author} {\bibfnamefont {A.}~\bibnamefont
  {Iorio}}\ and\ \bibinfo {author} {\bibfnamefont {P.}~\bibnamefont {Pais}},\
  }\href@noop {} {\bibfield  {journal} {\bibinfo  {journal} {Phys. Rev.}\
  }\textbf {\bibinfo {volume} {D92}},\ \bibinfo {pages} {125005} (\bibinfo
  {year} {2015})},\ \Eprint {http://arxiv.org/abs/hep-th/1508.00926}
  {hep-th/1508.00926} \BibitemShut {NoStop}%
\bibitem [{\citenamefont {Das}\ and\ \citenamefont
  {Vagenas}(2008)}]{DasVagenas}%
  \BibitemOpen
  \bibfield  {author} {\bibinfo {author} {\bibfnamefont {S.}~\bibnamefont
  {Das}}\ and\ \bibinfo {author} {\bibfnamefont {E.~C.}\ \bibnamefont
  {Vagenas}},\ }\href@noop {} {\bibfield  {journal} {\bibinfo  {journal} {Phys.
  Rev. Lett.}\ }\textbf {\bibinfo {volume} {101}},\ \bibinfo {pages} {221301}
  (\bibinfo {year} {2008})}\BibitemShut {NoStop}%
\bibitem [{\citenamefont {Ali}\ \emph {et~al.}(2009)\citenamefont {Ali},
  \citenamefont {Das},\ and\ \citenamefont {Vagenas}}]{AhmedAli1}%
  \BibitemOpen
  \bibfield  {author} {\bibinfo {author} {\bibfnamefont {A.~F.}\ \bibnamefont
  {Ali}}, \bibinfo {author} {\bibfnamefont {S.}~\bibnamefont {Das}}, \ and\
  \bibinfo {author} {\bibfnamefont {E.~C.}\ \bibnamefont {Vagenas}},\
  }\href@noop {} {\bibfield  {journal} {\bibinfo  {journal} {Phys. Lett.}\
  }\textbf {\bibinfo {volume} {B678}},\ \bibinfo {pages} {497} (\bibinfo {year}
  {2009})}\BibitemShut {NoStop}%
\bibitem [{\citenamefont {Das}\ \emph {et~al.}(2010)\citenamefont {Das},
  \citenamefont {Vagenas},\ and\ \citenamefont {Ali}}]{AhmedAli2}%
  \BibitemOpen
  \bibfield  {author} {\bibinfo {author} {\bibfnamefont {S.}~\bibnamefont
  {Das}}, \bibinfo {author} {\bibfnamefont {E.~C.}\ \bibnamefont {Vagenas}}, \
  and\ \bibinfo {author} {\bibfnamefont {A.~F.}\ \bibnamefont {Ali}},\
  }\href@noop {} {\bibfield  {journal} {\bibinfo  {journal} {Phys. Lett.}\
  }\textbf {\bibinfo {volume} {B690}},\ \bibinfo {pages} {407} (\bibinfo {year}
  {2010})}\BibitemShut {NoStop}%
\bibitem [{\citenamefont {Ali}\ \emph {et~al.}(2011)\citenamefont {Ali},
  \citenamefont {Das},\ and\ \citenamefont {Vagenas}}]{AhmedAli3}%
  \BibitemOpen
  \bibfield  {author} {\bibinfo {author} {\bibfnamefont {A.~F.}\ \bibnamefont
  {Ali}}, \bibinfo {author} {\bibfnamefont {S.}~\bibnamefont {Das}}, \ and\
  \bibinfo {author} {\bibfnamefont {E.~C.}\ \bibnamefont {Vagenas}},\
  }\href@noop {} {\bibfield  {journal} {\bibinfo  {journal} {Phys. Rev. D}\
  }\textbf {\bibinfo {volume} {84}},\ \bibinfo {pages} {044013} (\bibinfo
  {year} {2011})}\BibitemShut {NoStop}%
\bibitem [{\citenamefont {Maggiore}(1993)}]{Maggiore}%
  \BibitemOpen
  \bibfield  {author} {\bibinfo {author} {\bibfnamefont {M.}~\bibnamefont
  {Maggiore}},\ }\href@noop {} {\bibfield  {journal} {\bibinfo  {journal}
  {Phys. Lett.}\ }\textbf {\bibinfo {volume} {B304}},\ \bibinfo {pages} {65}
  (\bibinfo {year} {1993})}\BibitemShut {NoStop}%
\bibitem [{\citenamefont {Iorio}(2011)}]{weylgraphene}%
  \BibitemOpen
  \bibfield  {author} {\bibinfo {author} {\bibfnamefont {A.}~\bibnamefont
  {Iorio}},\ }\href {\doibase 10.1016/j.aop.2011.01.001} {\bibfield  {journal}
  {\bibinfo  {journal} {Ann. Phys. (N.Y.)}\ }\textbf {\bibinfo {volume}
  {326}},\ \bibinfo {pages} {1334} (\bibinfo {year} {2011})},\ \Eprint
  {http://arxiv.org/abs/hep-th/1007.5012} {hep-th/1007.5012} \BibitemShut
  {NoStop}%
\bibitem [{\citenamefont {Iorio}\ and\ \citenamefont
  {Lambiase}(2012)}]{hawkinggrapheneplb}%
  \BibitemOpen
  \bibfield  {author} {\bibinfo {author} {\bibfnamefont {A.}~\bibnamefont
  {Iorio}}\ and\ \bibinfo {author} {\bibfnamefont {G.}~\bibnamefont
  {Lambiase}},\ }\href@noop {} {\bibfield  {journal} {\bibinfo  {journal}
  {Phys. Lett.}\ }\textbf {\bibinfo {volume} {B716}},\ \bibinfo {pages} {334}
  (\bibinfo {year} {2012})},\ \Eprint
  {http://arxiv.org/abs/cond-mat.mtrl-sci/1108.2340}
  {cond-mat.mtrl-sci/1108.2340} \BibitemShut {NoStop}%
\bibitem [{\citenamefont {Iorio}\ and\ \citenamefont
  {Lambiase}(2014)}]{hawkinggrapheneprd}%
  \BibitemOpen
  \bibfield  {author} {\bibinfo {author} {\bibfnamefont {A.}~\bibnamefont
  {Iorio}}\ and\ \bibinfo {author} {\bibfnamefont {G.}~\bibnamefont
  {Lambiase}},\ }\href@noop {} {\bibfield  {journal} {\bibinfo  {journal}
  {Phys. Rev.}\ }\textbf {\bibinfo {volume} {D90}},\ \bibinfo {pages} {025006}
  (\bibinfo {year} {2014})},\ \Eprint {http://arxiv.org/abs/hep-th/1308.0265}
  {hep-th/1308.0265} \BibitemShut {NoStop}%
\bibitem [{\citenamefont {Iorio}\ and\ \citenamefont
  {Pais}(2018)}]{IORIO2018265}%
  \BibitemOpen
  \bibfield  {author} {\bibinfo {author} {\bibfnamefont {A.}~\bibnamefont
  {Iorio}}\ and\ \bibinfo {author} {\bibfnamefont {P.}~\bibnamefont {Pais}},\
  }\href@noop {} {\bibfield  {journal} {\bibinfo  {journal} {Ann. Phys.
  (N.Y.)}\ }\textbf {\bibinfo {volume} {398}},\ \bibinfo {pages} {265}
  (\bibinfo {year} {2018})}\BibitemShut {NoStop}%
\bibitem [{\citenamefont {Taioli}\ \emph {et~al.}(2016)\citenamefont {Taioli},
  \citenamefont {Gabbrielli}, \citenamefont {Simonucci}, \citenamefont
  {Pugno},\ and\ \citenamefont {Iorio}}]{icrystals}%
  \BibitemOpen
  \bibfield  {author} {\bibinfo {author} {\bibfnamefont {S.}~\bibnamefont
  {Taioli}}, \bibinfo {author} {\bibfnamefont {R.}~\bibnamefont {Gabbrielli}},
  \bibinfo {author} {\bibfnamefont {S.}~\bibnamefont {Simonucci}}, \bibinfo
  {author} {\bibfnamefont {N.~M.}\ \bibnamefont {Pugno}}, \ and\ \bibinfo
  {author} {\bibfnamefont {A.}~\bibnamefont {Iorio}},\ }\href@noop {}
  {\bibfield  {journal} {\bibinfo  {journal} {J. Phys. Condens. Matter}\
  }\textbf {\bibinfo {volume} {28}},\ \bibinfo {pages} {13LT01} (\bibinfo
  {year} {2016})}\BibitemShut {NoStop}%
\bibitem [{\citenamefont {Scardigli}\ \emph {et~al.}(2017)\citenamefont
  {Scardigli}, \citenamefont {Lambiase},\ and\ \citenamefont
  {Vagenas}}]{SCARDIGLI2017242}%
  \BibitemOpen
  \bibfield  {author} {\bibinfo {author} {\bibfnamefont {F.}~\bibnamefont
  {Scardigli}}, \bibinfo {author} {\bibfnamefont {G.}~\bibnamefont {Lambiase}},
  \ and\ \bibinfo {author} {\bibfnamefont {E.~C.}\ \bibnamefont {Vagenas}},\
  }\href@noop {} {\bibfield  {journal} {\bibinfo  {journal} {Phys. Lett. B}\
  }\textbf {\bibinfo {volume} {767}},\ \bibinfo {pages} {242} (\bibinfo {year}
  {2017})}\BibitemShut {NoStop}%
\bibitem [{\citenamefont {Snyder}(1947)}]{Snyder}%
  \BibitemOpen
  \bibfield  {author} {\bibinfo {author} {\bibfnamefont {H.~S.}\ \bibnamefont
  {Snyder}},\ }\href@noop {} {\bibfield  {journal} {\bibinfo  {journal} {Phys.
  Rev.}\ }\textbf {\bibinfo {volume} {71}},\ \bibinfo {pages} {38} (\bibinfo
  {year} {1947})}\BibitemShut {NoStop}%
\bibitem [{\citenamefont {Connes}(1994)}]{DBLP:books/daglib/0076876}%
  \BibitemOpen
  \bibfield  {author} {\bibinfo {author} {\bibfnamefont {A.}~\bibnamefont
  {Connes}},\ }\href@noop {} {\emph {\bibinfo {title} {Noncommutative
  geometry}}}\ (\bibinfo  {publisher} {Academic Press},\ \bibinfo {year}
  {1994})\BibitemShut {NoStop}%
\bibitem [{\citenamefont {Doplicher}\ \emph {et~al.}(1995)\citenamefont
  {Doplicher}, \citenamefont {Fredenhagen},\ and\ \citenamefont
  {Roberts}}]{Doplicher1995}%
  \BibitemOpen
  \bibfield  {author} {\bibinfo {author} {\bibfnamefont {S.}~\bibnamefont
  {Doplicher}}, \bibinfo {author} {\bibfnamefont {K.}~\bibnamefont
  {Fredenhagen}}, \ and\ \bibinfo {author} {\bibfnamefont {J.~E.}\ \bibnamefont
  {Roberts}},\ }\href@noop {} {\bibfield  {journal} {\bibinfo  {journal}
  {Commun. Math. Phys.}\ }\textbf {\bibinfo {volume} {172}},\ \bibinfo {pages}
  {187} (\bibinfo {year} {1995})}\BibitemShut {NoStop}%
\bibitem [{\citenamefont {Madore}\ \emph {et~al.}(2000)\citenamefont {Madore},
  \citenamefont {Schraml}, \citenamefont {Schupp},\ and\ \citenamefont
  {Wess}}]{Madore2000}%
  \BibitemOpen
  \bibfield  {author} {\bibinfo {author} {\bibfnamefont {J.}~\bibnamefont
  {Madore}}, \bibinfo {author} {\bibfnamefont {S.}~\bibnamefont {Schraml}},
  \bibinfo {author} {\bibfnamefont {P.}~\bibnamefont {Schupp}}, \ and\ \bibinfo
  {author} {\bibfnamefont {J.}~\bibnamefont {Wess}},\ }\href@noop {} {\bibfield
   {journal} {\bibinfo  {journal} {Eur. Phys. J.}\ }\textbf {\bibinfo {volume}
  {C16}},\ \bibinfo {pages} {161} (\bibinfo {year} {2000})},\ \Eprint
  {http://arxiv.org/abs/hep-th/0001203} {hep-th/0001203} \BibitemShut {NoStop}%
\bibitem [{\citenamefont {Seiberg}\ and\ \citenamefont
  {Witten}(1999)}]{Seiberg-Witten}%
  \BibitemOpen
  \bibfield  {author} {\bibinfo {author} {\bibfnamefont {N.}~\bibnamefont
  {Seiberg}}\ and\ \bibinfo {author} {\bibfnamefont {E.}~\bibnamefont
  {Witten}},\ }\href@noop {} {\bibfield  {journal} {\bibinfo  {journal} {JHEP}\
  }\textbf {\bibinfo {volume} {09}},\ \bibinfo {pages} {032} (\bibinfo {year}
  {1999})},\ \Eprint {http://arxiv.org/abs/hep-th/9908142} {hep-th/9908142}
  \BibitemShut {NoStop}%
\bibitem [{\citenamefont {Jackiw}(2002)}]{Jackiw2002}%
  \BibitemOpen
  \bibfield  {author} {\bibinfo {author} {\bibfnamefont {R.}~\bibnamefont
  {Jackiw}},\ }\href@noop {} {\bibfield  {journal} {\bibinfo  {journal} {Nucl.
  Phys. B - Proceedings Supplements}\ }\textbf {\bibinfo {volume} {108}},\
  \bibinfo {pages} {30} (\bibinfo {year} {2002})}\BibitemShut {NoStop}%
\bibitem [{\citenamefont {Guralnik}\ \emph {et~al.}(2001)\citenamefont
  {Guralnik}, \citenamefont {Jackiw}, \citenamefont {Pi},\ and\ \citenamefont
  {Polychronakos}}]{GURALNIK2001450}%
  \BibitemOpen
  \bibfield  {author} {\bibinfo {author} {\bibfnamefont {Z.}~\bibnamefont
  {Guralnik}}, \bibinfo {author} {\bibfnamefont {R.}~\bibnamefont {Jackiw}},
  \bibinfo {author} {\bibfnamefont {S.}~\bibnamefont {Pi}}, \ and\ \bibinfo
  {author} {\bibfnamefont {A.}~\bibnamefont {Polychronakos}},\ }\href@noop {}
  {\bibfield  {journal} {\bibinfo  {journal} {Phys. Lett. B}\ }\textbf
  {\bibinfo {volume} {517}},\ \bibinfo {pages} {450} (\bibinfo {year}
  {2001})}\BibitemShut {NoStop}%
\bibitem [{\citenamefont {Castorina}\ \emph {et~al.}(2004)\citenamefont
  {Castorina}, \citenamefont {Iorio},\ and\ \citenamefont
  {Zappal\`a}}]{PhysRevD.69.065008}%
  \BibitemOpen
  \bibfield  {author} {\bibinfo {author} {\bibfnamefont {P.}~\bibnamefont
  {Castorina}}, \bibinfo {author} {\bibfnamefont {A.}~\bibnamefont {Iorio}}, \
  and\ \bibinfo {author} {\bibfnamefont {D.}~\bibnamefont {Zappal\`a}},\
  }\href@noop {} {\bibfield  {journal} {\bibinfo  {journal} {Phys. Rev. D}\
  }\textbf {\bibinfo {volume} {69}},\ \bibinfo {pages} {065008} (\bibinfo
  {year} {2004})}\BibitemShut {NoStop}%
\bibitem [{\citenamefont {Balachandran}\ \emph {et~al.}(2006)\citenamefont
  {Balachandran}, \citenamefont {Pinzul},\ and\ \citenamefont
  {Qureshi}}]{BALACHANDRAN2006434}%
  \BibitemOpen
  \bibfield  {author} {\bibinfo {author} {\bibfnamefont {A.}~\bibnamefont
  {Balachandran}}, \bibinfo {author} {\bibfnamefont {A.}~\bibnamefont
  {Pinzul}}, \ and\ \bibinfo {author} {\bibfnamefont {B.}~\bibnamefont
  {Qureshi}},\ }\href@noop {} {\bibfield  {journal} {\bibinfo  {journal} {Phys.
  Lett. B}\ }\textbf {\bibinfo {volume} {634}},\ \bibinfo {pages} {434}
  (\bibinfo {year} {2006})}\BibitemShut {NoStop}%
\bibitem [{\citenamefont {Jiang}\ \emph {et~al.}(2007)\citenamefont {Jiang},
  \citenamefont {Henriksen}, \citenamefont {Tung}, \citenamefont {Wang},
  \citenamefont {Schwartz}, \citenamefont {Han}, \citenamefont {Kim},\ and\
  \citenamefont {Stormer}}]{PhysRevLett.98.197403}%
  \BibitemOpen
  \bibfield  {author} {\bibinfo {author} {\bibfnamefont {Z.}~\bibnamefont
  {Jiang}}, \bibinfo {author} {\bibfnamefont {E.~A.}\ \bibnamefont
  {Henriksen}}, \bibinfo {author} {\bibfnamefont {L.~C.}\ \bibnamefont {Tung}},
  \bibinfo {author} {\bibfnamefont {Y.-J.}\ \bibnamefont {Wang}}, \bibinfo
  {author} {\bibfnamefont {M.~E.}\ \bibnamefont {Schwartz}}, \bibinfo {author}
  {\bibfnamefont {M.~Y.}\ \bibnamefont {Han}}, \bibinfo {author} {\bibfnamefont
  {P.}~\bibnamefont {Kim}}, \ and\ \bibinfo {author} {\bibfnamefont {H.~L.}\
  \bibnamefont {Stormer}},\ }\href@noop {} {\bibfield  {journal} {\bibinfo
  {journal} {Phys. Rev. Lett.}\ }\textbf {\bibinfo {volume} {98}},\ \bibinfo
  {pages} {197403} (\bibinfo {year} {2007})}\BibitemShut {NoStop}%
\bibitem [{\citenamefont {Iorio}(2008)}]{PhysRevD.77.048701}%
  \BibitemOpen
  \bibfield  {author} {\bibinfo {author} {\bibfnamefont {A.}~\bibnamefont
  {Iorio}},\ }\href@noop {} {\bibfield  {journal} {\bibinfo  {journal} {Phys.
  Rev. D}\ }\textbf {\bibinfo {volume} {77}},\ \bibinfo {pages} {048701}
  (\bibinfo {year} {2008})}\BibitemShut {NoStop}%
\bibitem [{\citenamefont {Iorio}\ and\ \citenamefont
  {Sykora}(2002)}]{doi:10.1142/S0217751X02009874}%
  \BibitemOpen
  \bibfield  {author} {\bibinfo {author} {\bibfnamefont {A.}~\bibnamefont
  {Iorio}}\ and\ \bibinfo {author} {\bibfnamefont {T.}~\bibnamefont {Sykora}},\
  }\href@noop {} {\bibfield  {journal} {\bibinfo  {journal} {Int. J. Mod. Phys.
  A}\ }\textbf {\bibinfo {volume} {17}},\ \bibinfo {pages} {2369} (\bibinfo
  {year} {2002})},\ \Eprint {http://arxiv.org/abs/hep-th/0111049}
  {hep-th/0111049} \BibitemShut {NoStop}%
\bibitem [{\citenamefont {\'Alvarez-Gaum\'e}\ and\ \citenamefont
  {V\'azquez-Mozo}(2003)}]{ALVAREZGAUME2003293}%
  \BibitemOpen
  \bibfield  {author} {\bibinfo {author} {\bibfnamefont {L.}~\bibnamefont
  {\'Alvarez-Gaum\'e}}\ and\ \bibinfo {author} {\bibfnamefont {M.~A.}\
  \bibnamefont {V\'azquez-Mozo}},\ }\href@noop {} {\bibfield  {journal}
  {\bibinfo  {journal} {Nucl. Phys. B}\ }\textbf {\bibinfo {volume} {668}},\
  \bibinfo {pages} {293} (\bibinfo {year} {2003})}\BibitemShut {NoStop}%
\end{thebibliography}%

\end{document}